\documentclass[conference]{IEEEtran}
\IEEEoverridecommandlockouts
\usepackage{cite}
\usepackage{amsmath,amssymb,amsfonts}
\usepackage{algorithm}
\usepackage[noend]{algpseudocode}
\usepackage{graphicx}
\usepackage{textcomp}
\usepackage{xcolor}
\usepackage{subcaption}
\usepackage{authblk}
\usepackage{microtype}
\usepackage{empheq}

\usepackage{xurl}
\usepackage{hyperref}

\def\BibTeX{{\rm B\kern-.05em{\sc i\kern-.025em b}\kern-.08em
    T\kern-.1667em\lower.7ex\hbox{E}\kern-.125emX}}
\begin{document}

\title{Data-based optimal estimation of frequency bias: The case of Southwest Power Pool 
\thanks{Partial funding for this work is provided by the US National Science Foundation EAGER project \#2002570} }

\author[1]{Miroslav Kosani\'{c}}
\author[1]{Marija Ili\'{c}}
\author[2]{Daniel Baker} 
\author[2]{Harvey Scribner}
\author[2]{Casey Cathey}
\affil[1] {Massachusetts Institute of Technology, Cambridge, MA 02139; ilic@mit.edu; kosanic@mit.edu}
\affil[2]{Southwest  Power Pool, Little Rock, AR 72223; dbaker@spp.org;hscribner@spp.org;ccathey@spp.org}



\maketitle

\begin{abstract}
In this paper, we introduce   a method to optimally estimate time-varying frequency bias $\beta$.  Current industry practice is to assume that $\beta$ is changing only on annual basis. We suggest that this improved time-dependent bias estimate can be used to reduce the cost of frequency regulation needed to meet  industry standards requested by the North American Electric Reliability Corporation (NERC). Optimization of time-varying frequency bias is posed as a parameter estimation (calibration) problem  whose implementation utilizes  online  system  measurements.   It is further  shown how this result can be used to estimate intra-dispatch load deviations. This knowledge is needed to  estimate more accurately regulation reserve needed, and to therefore reduce overall regulation cost.  Methods can be introduced to give incentives to demand response to participate in frequency regulation. Overall,   we show  the importance of incorporating knowledge of physics-based models for data-enabled parameter estimation of physical systems. 
\end{abstract}


\section{Introduction}


An  AC interconnected electric power system is a complex system whose main function is to provide uninterrupted electricity service.  Therefore,   the system must be operated within  pre-specified frequency deviations around nominal frequency  by maintaining online balance of net generation and load. This is accomplished by ensuring that adequate resources are available to respond to power imbalances as they  happen.  Fast-responding power plants are usually used to do this in a feedback manner. The missed opportunity cost associated with scheduling more expensive polluting power plants to supply predictable system load has caused the cost of fast  power plants which regulate frequency in response to system load deviations  to be historically high \cite{hoffman}.
\subsection{Frequency stabilization  and regulation in today's industry}
Frequency stabilization and regulation are inherent functions of today's hierarchical control \cite{iliczab}. Primary  control of generator-turbine-governor (GTG) is a fast local control responding to the deviations of frequency  $\omega(t)$ from the frequency set point $\omega^{ref} [kTs]$ of their governors. Set points of  power plants participating in secondary control  comprise Automatic Generation Control (AGC) function which in an automated manner adjusts set points  $\omega^{ref} [kTs]$  on their governors so that the Balancing Authority (BA) level regulates frequency within the pre-specified limits.  Each area $I$ comprises several  generators $G_j^I$ participating   in the BA frequency regulation.  The AGC function operates in a feedback manner by responding to the total area imbalance known as the Area Control Error (ACE) on a minute-by-minute basis   and it does not require 
the system operator to make decisions in near real-time.  However,  as the generation mix is rapidly changing, the natural response of  each area $I$  varies over time since, as reviewed later in the paper,  the sensitivity of frequency deviations with respect to power imbalances depends on the GTG parameters of all power plants participating in AGC. Also, the ACE variations are  becoming more dynamic and higher in amplitude as they reflect the net load deviations from the predicted system load, which  routinely includes the effects of intermittent power generation and  BA  demand deviations from  their historic patterns.  All these changes have made it much more challenging to estimate the amount of regulation reserve, and this affects the overall cost of meeting frequency regulation  standards.
\begin{figure}
  \centering
  \includegraphics[scale=.5]{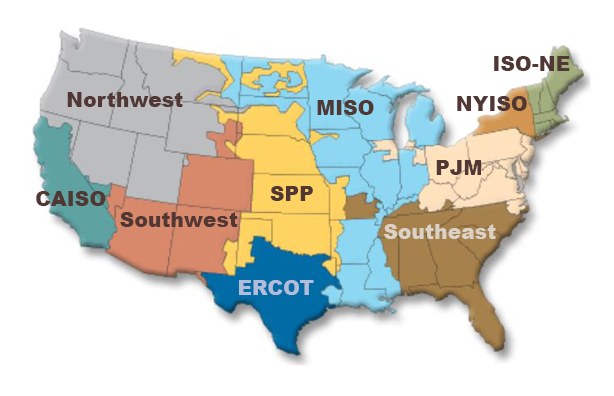}
  \caption{Map of North America independent system operators (ISOs)}
  \label{fig_spp}
\end{figure}
\subsection{Pricing regulation reserve in electricity markets}
Furthermore, in areas of the US interconnection in which generation is provided competitively through the   evolving electricity markets the Federal Energy Regulatory Commission (FERC), under order 2000 has defined ancillary services as a means to balance power  during both normal and abnormal conditions, the latter  caused by the large equipment failures.  By their definition, ancillary services maintain reliable operations of the interconnected transmission system. As it is stated on the FERC website ~\cite{ferc}, they encompass load following, reactive power-voltage regulation, system protective services, loss compensation service, system control, load dispatch services, and energy imbalance services.  The cost of balancing power during normal conditions, well understood in the regulated industry as the AGC cost, is included as part of the ancillary service cost. In the past, ACG costs were not considered to have a major effect on the market electricity prices.  This is rapidly changing as this cost is increasing, and, therefore,  it is also very important to provide better estimates of the regulation reserves typically purchased in day-ahead markets.  As a result, market mechanisms for frequency regulation have taken on a new importance and the problem is undergoing its renaissance \cite{pxfc}.  

\subsection{Growing concerns}
System operators in all BAs in the US, regulated or market supported, are concerned with the growing influx of intermittent power and its effects on the regulation reserves. Shown in Figure \ref{fig_spp}  
is SPP BA whose measurement data we are using in this paper. SPP has had a huge growth in both wind and utility-scale power generation. Estimating net load deviations in the SPP area is important for planning and utilizing regulation reserves as these types of energy resources are  deployed in larger amounts.  
 To reduce the overall cost of electricity service, it is becoming increasingly important to estimate net system load deviations around their predicted patterns much more accurately.   It is also becoming more important to provide incentives to demand response as a means of balancing these minute-by-minute fast deviations \cite{black,demand}. 

Notably, the SPP net load volatility has increased since 2016 and is expected to further increase. Penetration of renewables, primarily wind generation, increased these changes in net load but also transmission congestion (Oklahoma has one of the single-largest wind farms in North America)  as can be seen from quarterly presentation in 2022 ~\cite{sppmmu}. SPP attempted to address the first issue as a way of keeping reliability through offering ramp products to systematically pre-position resources with ramp capability to manage net load variations and uncertainties ~\cite{sppferc} and provide transparent price signals to incentivize resource flexibility and future economic investment. Since then, there is an energy price increase ~\cite{sppmmu}.
\subsection{Paper organization}
This paper is motivated by these overall  growing concerns about the ability to regulate frequency at a reasonable cost as described  above. The basic premise is that by systematic data mining, it  is possible to estimate both time-varying  natural response of a BA, SPP in particular, to consequently estimate more accurately  minute-by-minute load deviations $\Delta P_L[kTs]$ and assist system operators and markets in purchasing adequate regulation reserve.  Section \ref{agc} sets an overview of AGC and Section \ref{estimation} sets the  physics-based model structure essential for effective estimation of natural BA response. The problem is posed as an inverse optimization problem of BA droop characteristic  deviation from its physics-based model.  Data is used to compute these deviations and to estimate BA droop characteristic, in particular BA frequency bias. In Section ~\ref{results} SPP data is used to demonstrate the time-varying frequency bias estimates around the constant SPP value given to us. Estimates of the  potential saving on the amount of regulation reserve needed and the resulting  regulation reserve cost are briefly described.   Finally, in Section  ~\ref{money}  we discuss and conclude  with open questions for future research.

\section{Automatic generation control}
\label{agc}
Automatic Generation Control (AGC) in the US interconnected power system or Load Frequency Control (LFC) in Europe,  have been examples of the most ingenious  large-scale feedback control schemes of complex man-made dynamical systems.  They have worked amazingly well, despite their simplicity, and have been the key to regulating interconnection frequency. The main objective  of each Control Area (CA) comprising  a subsystem within a large-scale interconnected system,  has been to  implement frequency regulation by re-setting the set points of governor controllers $\omega^{ref}[kT_s]$ of  power plants participating in AGC  so that the ACE is compensated by their supplemental generation $P^{reg}[kT_s]$. This is done automatically  in a feedback manner on a minute-by-minute basis $[kT_s]$  around feed-forward tertiary level  generation scheduling in between dispatch times $[kT_t]$.  Historically  AGC was implemented using mainly  fast power plants most  suited to produce power fast and contribute their share of regulating power to cancel ACE of a CA.  More recently,  the US Control Areas (CAs) have merged into larger Balancing Authorities (BAs) shown  in Figure \ref{fig_spp}. 
NERC BAL-003-1.1 standard requires that AGC for the interconnected system should regulate  frequency within the pre-specified  deviation limits of  $\pm 0.036 Hz$ around the nominal $60 Hz$ frequency. This is done by each BA in a distributed manner  according to its  natural response published by NERC \cite{nerc2017}. Electricity markets require that regulation reserves are price-based and, as such, 
the frequency regulation process has become more complex. 
\subsection{Physics-based AGC  model}
\begin{figure}[h!]
  \centering
  \includegraphics[scale=.4]{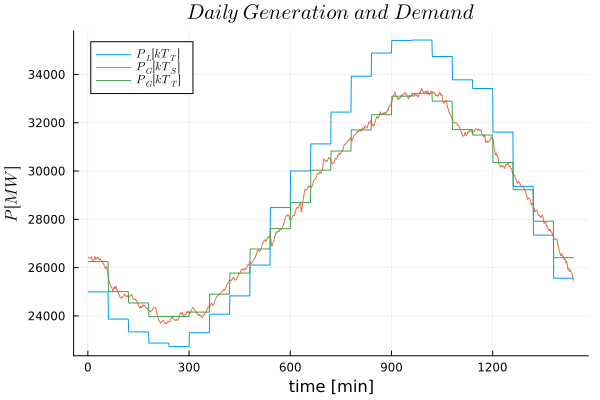}
  \caption{SPP Generation and demand for 2017-10-14}
  \label{fig:gendem}
\end{figure}

For completeness, we  briefly review the physics-based model used for implementing today's AGC. Tertiary level feed-forward scheduling of generation $ \hat P_G[kT_t]$ is done hourly at times $[kT_t]$ to supply predictable component of system load $\hat P_l[kT_t]$ and the 
pre-agreed on Net Scheduled Interchange (NSI) with the neighboring BAs so that 
\begin{align}
    \hat P_G[kTt] = NSI[kTt] + \hat P_L[kTt]
    \label{eq:hatpg}
\end{align}
Closer to real-time,  power plants participating in frequency regulation $P^{reg}[kT_s]$ adjust their governor set points $\omega^{ref}[kT_s]$ to compensate actual load $P_L[kT_s]$ and Net Actual Interchange $NAI[kT_s]$ so that the sum of generation scheduled in a feed-forward manner and regulation power balance actual NAI and actual load each $[kT_s]$ time interval, typically minute-by-minute. 
\begin{align}
    P_G[kTt] + P_G^{reg}[kTs] = NAI[kTs] + P_L[kTs]
    \label{eq:pg}
\end{align}
  Shown in Fig. \ref{fig:gendem} is  the mismatch between predicted demand and generated and predicted generation for the SPP system. This mismatch  is generally caused by:
  \begin{enumerate}
      \item not knowing the actual load $P_L[kT_s]$ SPP incorrect prediction/measurement of the load as can be seen  in Fig. \ref{fig:gendem}), violating the  power balance stated in Eqn. (\ref{eq:pg})
      \item and/or by the deviations of net tie-line flow interchange $\Delta F[kT_s]$  defined as 
          \begin{equation}
              \Delta F[kT_s] = NAI[kT_s] -NSI [kT_t]
              \label{eqn_flowdev}
          \end{equation}
  \end{enumerate} 

Frequency deviation from nominal is generally affected by the mismatch of production and generation, as well as by the net interchange deviations of $NAI[kT_s]$ from the net scheduled interchange $NSI[kT_t]$. Combining Eqn. (\ref{eq:hatpg}) and Eqn. (\ref{eq:pg}) and decomposing  power imbalance into:
\begin{enumerate}
    \item imbalance created by the deviations of  actual load $P_L[kT_s]$ from the predicted load $\hat P_L[kT_t]$ denoted as  $ACE_{f}[kT_s]$
    \item power imbalance  $\Delta F[kT_s]$  between   $NSI[kT_s]$ and $NAI[kT_S]$,   denoted as $ACE_{interchange}[kT_s]$, we obtain: 
        \begin{equation}
            ACE [kT_s] = ACE_{f} [kT_s]+ \Delta F[kTs]
            \label{eq:acesum}
        \end{equation}
        where $ACE_f[kT_s]$ is the frequency part of the ACE and $\Delta F[kT_s]$  is the  interchange part of the ACE.
\end{enumerate}   

  To compensate for  internal  load deviations, nonzero generation by  the power plants participating in regulation is provided in a feedback manner  by adjusting  the  set points of  their governors $\omega^{ref}[kT_s]$ as
\begin{equation}
P_G^{reg[kT_s]} = -10 b \Delta f[kT_s] = \beta \Delta f[kT_s]
\label{PGreg}
\end{equation}
Determining by how much  to change the set point of governors  so that  the  $P_G^{reg}[kT_s]$  is produced for frequency regulation purposes  requires to know  $\beta$ of the BA. In what follows we briefly summarize the physical interpretation of this coefficient known as the BA frequency bias. The actual amount of $P_G^{ref}[kT_s]$ that must be compensated by the BA is determined by the very hard to know $\Delta P_L[kT_s]$. The amount of regulation reserve needed to produce this power cannot be known in a feed-forward way, say day-ahead or hour-ahead.   However, by measuring frequency deviations $\Delta f[kT_s]$ and using frequency bias $\beta$ one can estimate  at least its  hourly bounds. This is one of the main reasons we seek a method for estimating time-varying $\beta[kT_t]$, so that the regulation reserve is scheduled ahead of time and used in a feedback manner in between dispatch time intervals $[kT_t]$.

BA loads are much more dynamic and harder  and make it difficult to perform  accurate estimation of frequency bias $\beta$ in systems   with a  variety of  intermittent  distributed energy resources (DERs) such as wind and solar power plants, as well as the price-responsive demand.  $ACE_{frequency}$ portion of $ACE$ is complex to obtain and control under the presently made assumptions.  As AGC input is ACE there is the need for accurate frequency bias  $\beta$ which reflects CA droop, namely the sensitivity of the frequency deviations  to deviations in power imbalances, $\beta = 10 b$, units of $\beta$   in
$\frac{MW}{Hz}$ or units of $b$ in  $\frac{MW}{.1Hz}$.  

To compute $ACE_{f}$ portion of $ACE$ the  knowledge of $\beta$ is quite critical. Notably,  the need for supplemental control is significantly smaller than it would be if it were not for the natural self-regulation by  most of the loads. The load power consumption is greatly dependent on  both voltage and frequency deviations, but these are not typically modeled \cite{black}.  In the changing industry with price-responsive demand it becomes quite important to account for this effect, a major open problem.  The intermittent power outputs also require much more dynamic dispatch of units than in the past, and this directly affects the frequency bias of the CA.

\section{Optimal Frequency Bias Estimation Method}
\label{estimation}
\label{sec:bias}
In this section we first establish the basic structure of the BA model needed to perform data-enabled parameter estimation, and then propose an otpimization method which draws on this structure. 

\subsection{Basic structure of the BA mode}
Basic structure of the BA model is obtained by utilizing the droop characteristics of individual regulating power plants and then aggregating them without taking into account the transmission grid. This is the most common assumption made  for modeling bulk power system; for more detailed  models, see \cite{shell,eagc}. 
To explain, we first summarize structure of a single generator droop and then derive an aggregate droop comprising all regulating power plants. 
\subsubsection{Droop characteristic of a single regulating power plant}
We  derive governor-turbine-generator (GTG) power plant droop  by combining  the dynamical equations  of a generator and turbine, with  the  governor dynamical equation \cite{shell,iliczab}:
\begin{align}
\dot \theta_G   & =  \omega_0 \omega_G
\label{eq:teta}\\
J \dot \omega_G + D \omega_G  & =  P_T + e_T a  -P_G 
\label{eq:omega}\\
T_u \dot P_T &  =  - P_T + K_t a
\label{eq:PT}\\
T_a \dot a  & =   - r  a - \omega_G + \omega_G^{ref}
\label{eq:valve}
\end{align}
The mechanical power $P_T$ applied to the turbine-rotor shaft is controlled by the valve position $a$  according to Eqn. (\ref{eq:PT}). Constant $T_u$ is the turbine time constant while $K_t$ is the control gain. Valve position $a$ changes in feedback manner by responding to the deviations of measured frequency at time $t$ from the set point of the governor at time $[kTs]$ $ \Delta \omega = (\omega (t) -\omega^{ref}[kT_s])$, according to Eqn.(\ref{eq:valve}). The dynamics of frequency deviation $\omega_G$  from nominal frequency $\omega_0$ is determined by the imbalances between the mechanical power $P_T$ and electrical power generated $P_G$, and is further damped because of damping  $D$. Different power plant inertia $J$ result in different frequency dynamics.  The governor set  point is changed in response to tertiary level economic dispatch commands each $[kT_t]$, and it is re-adjusted  on power plants participating in AGC each $[kT_s]$. The primary control of the governor is tuned so that valve dynamics settles in response to fast small fluctuations to $\omega_G^{ref}[kT_s]$. Parameters in this GTG dynamical model are derived by linearizing more complex nonlinear power plant dynamics \cite{iliczab}. Important for understanding droops is  the definition of parameter $r$ in  Eqn.(\ref{eq:valve}), which determines how well is valve control tuned.  If done right, the fast dynamics evolving at continuous time settles and at times $[kT_s]$ results in the well-known droop of the GTG as follows \cite{iliczab}
\begin{equation}
\omega_G[k] = (1- \sigma D) \omega_G ^{ref}[k] - \sigma P_G [k]
\label{eq:droop}    
\end{equation}
  where GTG droop constant is defined as 
  \begin{equation}
    \sigma = \frac{\delta \omega_G[k]}{\delta P_G[k]}
    \label{e:droopconstant}
  \end{equation}
 
Understanding this three-way relationship given in Eqn. (\ref{eq:droop}) is critical for  understanding how frequency is controlled in today's bulk power systems by the conventional AGC. 
\subsection*{ Frequency bias and regulation  of a single  BA}
It can be seen from Eqn. (\ref{eq:droop})  that  the steady-state frequency deviation of generator $j$ in area $I$   $\omega_{G,j}^I = - \sigma_{G,j}^I P_{G,j}^I$  when the set point  $\omega _{G,j}^{I,ref}$ is not changed.  Since the steady-state frequency is the same in the entire area $\omega^I$, it can be   derived by summing the droop characteristics of all regulating units $j \in I$ a single aggregate area droop characteristic  of the form 
\begin{align}
 \omega ^I[kTs] & = \alpha^I [kTs]  \omega ^{I,ref}[kTs] - \sigma^I[kTs]   P^I [kTs]
  \label{eq:droopI}
\end{align}

where  the aggregate natural response is the sum of the natural responses of all units and $\alpha^I[kT_s] = (1-\sigma^I D^I)[kT_s]$
\begin{align}
\beta ^I[kT_s] &=  \Sigma _{j \in I} \frac{1}{\sigma_j}
\label{aggbeta}
\end{align}
and $\alpha = $
The aggregate reference frequency $\omega ^{I,ref}$ satisfies 
\begin{align}
(1-\sigma^I D^I) \omega  ^{I,ref} & =  \Sigma_{j \in I} (1-\sigma_j D_j)  \omega_j^{ref} 
\end{align}

\subsection*{AGC in two interconnected BAs}
Consider now an interconnected system comprising multiple BAs, as shown in Figure \ref{fig_spp}.   Equation  (\ref{eq:droopI}) can be -written for any area $S$ as 
\begin{align}
  \omega_S[kTs] &  =  \alpha[kTs] \omega_S ^{ref}[kTs] - \sigma[kTs]   P_S [kTs]
 \label{eq:droopM}
 \end{align}
However, in multi-area interconnected system an additional complexity arises because the net power generation $P^S[kTs]$  of any area $S$ must balance both its own load $P_L^S[kT_s]$ and the actual  net interchange with the neighbouring BAs $NAI^S[kT_s]$. This leads to the $ACE^S[kT_s]$ which can be decomposed into two components, one caused by the internal load deviations and the other caused by the deviations in net tie line flows from schedules. As  result,  for the case of two-control areas one obtains
\begin{equation}
ACE^S[kT_s] = ACE_f ^S[kT_s]  + ACE_{interchange}^S[kT_s]
\label{eq:SCEtwoareas}
\end{equation}
where  $ACE_{interchange}^S [kT_s] = \Delta F^S [kT_s]$, for any  $S$  comprising  multi-area inteconnected system. 

\subsection{Area frequency bias estimation method}
Eqn. (\ref{eq:droopM}) represents an aggregate model of single BA  with a very distinct structure.  We  formulate next inverse optimization problem. Using provided SPP data, natural response of area $\beta^I [kT_t]$  is estimated, by viewing it as a slower varying parameter  at times $kT_t$ given much faster state frequency state measurements $\omega^I[kT_s]$ and area  generation $P^I_G[kT_s]$. Frequency is related to angular frequency as $w = 2\pi f$.
Eq. (~\ref{eq:droopM}) can be posed as the problem of parameter calibration through multiple linear regression. Coefficients of interest $\alpha_S[kTs], \sigma_S[kTs]$ are obtained through minimization of the cost function
\begin{align}
    \resizebox{0.85\hsize}{!}{$J(\alpha_S[k], \beta_S[k]) = \sum_k{(f_S[k] - \alpha_S[k] f_S ^{ref}[k] + \sigma_S[k]   P_S [k]})^2$}
    \label{eq:criteria}
\end{align}
This formulation of regression objective function, where optimization is performed with respect to L2 norm, is called ordinary least squares (OLS) multiple regression. OLS has a closed form solution
\begin{align}
    \begin{bmatrix}
        \alpha \\
        \sigma
    \end{bmatrix} 
    =
    \left(\begin{bmatrix}
        f_s^{ref}[k] \\
        -P_S[k]
    \end{bmatrix}
    \begin{bmatrix}
        f_s^{ref}[k] \\
        -P_S[k]
    \end{bmatrix}^{T} \right)^{-1}
    \begin{bmatrix}
        f_s^{ref}[k] \\
        -P_S[k]
    \end{bmatrix} f_s[k]
    \label{eq:ols}
\end{align}
and through solving closed form, we obtain coefficients $\sigma$ and $\alpha$.


Decision variables $\alpha$ and $\sigma$ are changing, so minimizing over the whole day (24h or 1440 minutes) as in (Eq. \ref{eq:criteria})  we use (Eq. \ref{eq:ols}, and observe changes of variables of interest, minute by minute in the next section. 




\section{Numerical results using SPP system}
\label{results}
In the data provided to us, we were given the value of frequency bias used for the adjustment of frequency part of the ACE. This value is constant, computed annually and provided by NERC. Thus, its crude and constant value (for the year of 2017, it was decided to be $\beta = 409 MW/0.1Hz$) cannot be used for estimation of the load $P_L[kTs]$ at the timescale of the secondary frequency control. Results in this section show how frequency bias value changes through the day. We show incentive for adoption of time dependent frequency bias and its use for estimation of load deviations. Lastly, we show influence of interchange portion of ACE on the scheduling of reserves. Cooperative control and right information exchange between BAs, as we argue, would thus help in reducing costs of regulation reserves and give us better understanding of how much storage will be needed.

\subsection{Data provided by the system operator}
SPP provided us with the data summarized in the table \ref{table:1}. This data is sampled on a minute basis.
\begin{table}[h!]
    \centering
    \begin{tabular}{ |p{2cm}||p{5cm}|  }
     \hline
     \multicolumn{2}{|c|}{Summary of available data} \\
     \hline
     Data & Short description\\
     \hline
     $\Delta T[min]$   & interchange part of ACE in MW\\
     $ACE_{f}[min]$  & frequency part of ACE in MW\\
     $f_{ref}[min]$ &  scheduled frequency in Hz\\
     $f[min]$ &  measured frequency in Hz\\
     $P_{G}[min]$ & actual generation of the system\\
     $NAI[min]$ & net actual interchange\\
     \hline
    \end{tabular}
\caption{Provided data}
\label{table:1}
\end{table}

\subsection{Optimal frequency bias daily values}

\begin{figure}[h!]
\centering
\includegraphics[width=0.4\textwidth]{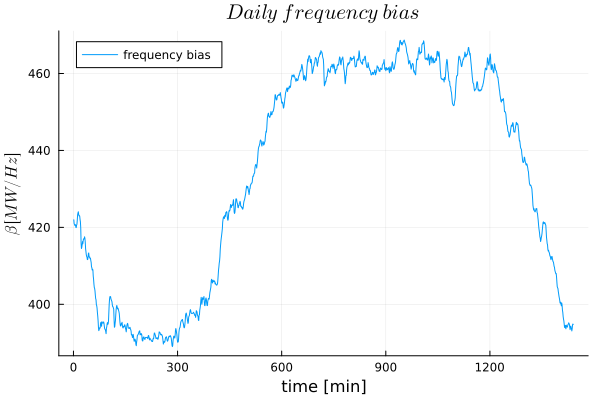}
\caption{\label{fig:fb1}  Optimal bias for 10-14-2017 on a one minute basis }
\end{figure}

\begin{figure}[h!]
\centering
\includegraphics[width=0.4\textwidth]{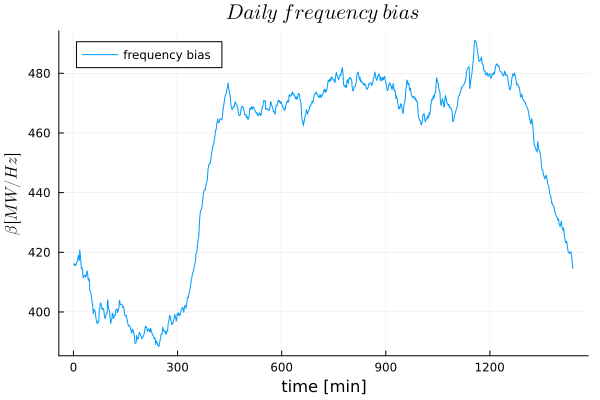}
\caption{\label{fig:fb2}  Optimal bias for 10-24-2017 on a one minute basis }
\end{figure}

\begin{figure}[h!]
\centering
\includegraphics[width=0.4\textwidth]{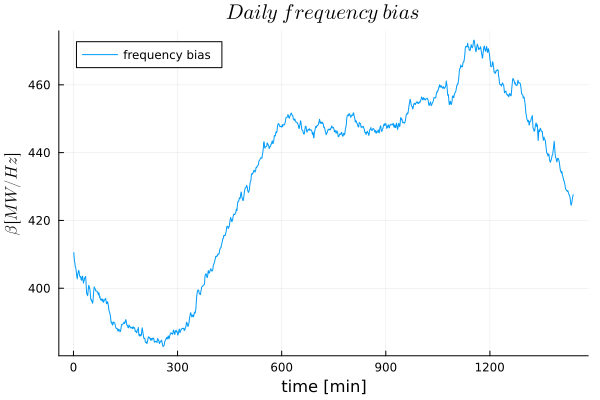}
\caption{\label{fig:fb3}  Optimal bias for 10-26-2017 on a one minute basis }
\end{figure}

The droop (Eq. ~\ref{eq:droop}) has such a structure that poses itself naturally for calibration of its unknown parameters through different data methods. For our purposes, we've used OLS multiple linear regression. Optimal value of frequency bias indicates that this value changes through the day as can be seen on the (Fig. ~\ref{fig:fb1},  ~\ref{fig:fb2} and ~\ref{fig:fb3}). Opposite to the constant value of $\beta$ we were provided by SPP, we observe that during the night, between 0-300 minutes (midnight-5am) frequency bias has a dip. From 5am onward, there is a rise, which settles around noon until 8pm, when trend of $\beta$ goes down. Between (Fig. ~\ref{fig:fb1},  ~\ref{fig:fb2} and ~\ref{fig:fb3}) we see slight differences (e.g. morning rising slope), likely related to (Fig. ~\ref{fig:fb1} being Saturday, and later two Tuesday and Thursday.

\section{Impact on secondary regulation reserves annual cost}
\label{money}

Using the estimate of frequency bias, learned through the knowledge of past generation and frequency curves, one can obtain estimate of load deviations. Up to our knowledge this is one and only way. 
\begin{equation}
    \Delta P_L[kTs] = -10\beta[kTs] \Delta f[kTs]
\end{equation}


\begin{figure}
  \centering
  \includegraphics[scale=.4]{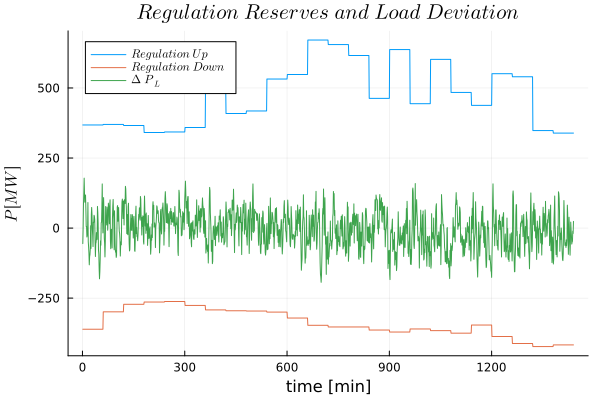}
  \caption{Regulation reserves boundaries compared to estimated load deviation on secondary time scale for 2017-10-14}
  \label{fig:regupdndpl}
\end{figure}

\begin{figure}
  \centering
  \includegraphics[scale=.4]{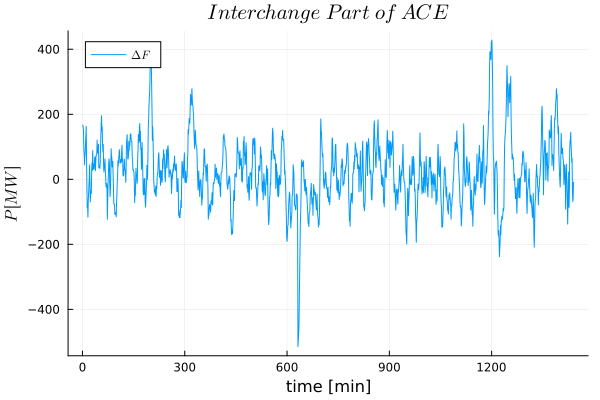}
  \caption{Interchange $\Delta F$ part of ACE for the 2017-10-14}
  \label{fig:interchange}
\end{figure}

\begin{figure}
  \centering
  \includegraphics[scale=.4]{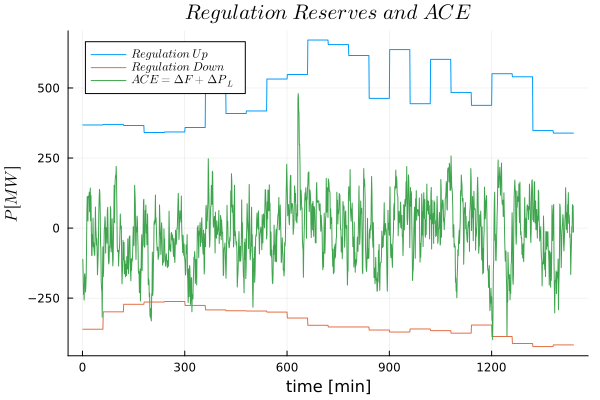}
  \caption{Regulation reserves boundaries compared to ACE on secondary time scale for 2017-10-14}
  \label{fig:regupdnace}
\end{figure}

Regulation-up and regulation-down have the highest market clearing
prices. We observe on (Fig. ~\ref{fig:regupdnace}) how spill of interchange power through $\Delta F$ part of ACE prevents tighter bounds of the day ahead regulation reserves. If $NAI$ matches $NSI$ (Fig. ~\ref{fig:interchange}) tighter bounds would then mean decrease of scheduled reserves by $200 W$ on average per minute, which translates to roughly $\$2.88 \cdot 10^6$ assuming average cost \cite{SPPMarket2017}, of the up $\$10$ and down $\$10$ regulating reserves. As can be seen in state of the market SPP annual reports trend of average regulation prices has been slightly increasing, but the the real benefits of the cooperative BAs control is not for the whole year, but for the spring and autumn months, when the wind production is high and the load is low. In these months (e.g. March, April, May and September, October, November), there is $33\%$ increase in regulation up reserves. Tighter bounds, possible through information exchange between BAs and decentralized control could potentially land in annual $\$259.2 \cdot 10^6$ of savings in previously mentioned months.







\section{Conclusion}
\label{conclusion}
Dynamic changes of frequency bias are happening throughout the day. Using the droop three-way relationship, we formulate optimization problem, which provides us with the optimal frequency bias value, along with the damping coefficient of the system. Both of these quantities change with time. Current practice where droop is constant should be revised, as there are clear economic incentives for change and participation of demand-response. Understanding wind penetration influence on regulation price reserves during the year incentivizes for cooperation between BAs so that tighter bounds on regulation reserves schedules can be imposed.  In other words, depending on what you monitor, estimate, how information exchange is performed and if BAs help each other,  one can integrate demand-response and in the process gain value. 

Future research would go in the direction of better understanding cooperation, and information exchange between BAs and how taking into account for electrical distances inside BAs influences the estimation of the load.




\bibliographystyle{plain}
\bibliography{text/bibliography}

\end{document}